\documentclass[showpacs,preprintnumbers,amsmath,amssymb]{revtex4}

\usepackage{graphicx}

\date{\today}

\begin{document}



\title{Dynamical aspects of quantum entanglement for weakly coupled 
kicked tops}

\author{Hiroshi Fujisaki}\email{fujisaki@ims.ac.jp}
\affiliation{%
Department of Theoretical Studies, Institute for Molecular Science,
Myodaiji, Okazaki, 444-8585, Japan 
}%

\author{Takayuki Miyadera}\email{Tmdella@aol.com}
\affiliation{%
Department of Information Sciences, 
Tokyo University of Science, Noda City, 
278-8510, Japan
}%

\author{Atushi Tanaka}\email{tanaka@phys.metro-u.ac.jp}
\affiliation{%
Department of Physics, Tokyo Metropolitan University,
Minami-Osawa, Hachioji 192-0397, Japan 
}%

\begin{abstract}
We investigate 
how the dynamical production of 
quantum entanglement for weakly coupled, composite quantum systems 
is influenced by the chaotic dynamics of the corresponding 
classical system, using coupled kicked tops.
The linear entropy for the subsystem (a kicked top) is employed  
as a measure of entanglement.
A perturbative formula for the entanglement production rate 
is derived. The formula contains a correlation function
that can be evaluated only from the information of uncoupled tops.
Using this expression and the assumption that the 
correlation function decays exponentially 
which is plausible for chaotic tops, 
it is shown that {\it the increment of the strength of chaos does not 
enhance the production rate of entanglement} 
when the coupling is 
weak enough and the subsystems (kicked tops) are strongly chaotic.
The result is confirmed by numerical experiments.
The perturbative approach is also applied to a weakly chaotic
region, where tori and chaotic sea coexist in the corresponding
classical phase space, to reexamine a recent numerical
study that suggests an intimate relationship between the linear
stability of the corresponding classical trajectory and the
entanglement production rate.

\end{abstract}


\maketitle

\section{Introduction}

Quantum entanglement 
(in short, entanglement)
in composite systems 
has been discussed as a paradoxical issue~\cite{WZ82},
but it is experimentally confirmed, 
and utilized in quantum information processing~\cite{NC00}.
Although the definition of entanglement itself 
is not of dynamical nature, 
entangled states are often generated 
{\it dynamically}~\cite{ZP94,decoherence}.
That is, even if subsystems are not entangled initially, 
the interaction between them produces 
entanglement in the system as time elapses.
It is easily expected that the dynamical production of quantum
entanglement heavily depends on the qualitative nature of dynamics. 
An important qualitative distinction of quantum dynamics is provided by
the corresponding classical dynamics, whether it is regular or
chaotic, as is well known in the literature of 
``quantum chaos''~\cite{Guztwiller}.
Hence, there have been investigations to answer the following
question~\cite{Adachi92,Tanaka96,SKO96,FNP98,Lakshminarayan01}:
{\it Does the production of entanglement depend on 
whether the corresponding classical system is chaotic or regular?}
The authors of 
Refs.~\cite{Adachi92,Tanaka96,SKO96,FNP98,Lakshminarayan01}
concluded that chaotic systems 
tend to produce larger entanglement than 
the regular systems in general
(exceptions are shown in Refs.~\cite{Tanaka96,AFNP99}).
Accordingly, it is natural to raise the next question:
{\it For chaotic systems, does the strength of chaos increase 
the degree of entanglement?}
This issue was first investigated by Miller and Sarkar~\cite{MS99}.
They employed coupled kicked tops (CKTs) 
as their model system, 
and numerically found 
that the von Neumann entropy 
of the subsystem linearly depends on the sum of 
positive (finite-time) Lyapunov exponents
of the corresponding classical systems.
To authors' knowledge, however,
there has been no theoretical explanation for their result. 

In this paper we examine the same system (CKTs) 
to elucidate the mechanism of the entanglement production. 
In particular,  
we clarify how dynamical aspects of  
quantum entanglement for CKTs 
are affected by the chaotic dynamics of the corresponding 
classical CKTs when the coupling is weak {\it and}
the chaos is strong enough. This limiting case 
should be examined first, and has not been fully 
investigated in previous studies.

This paper is organized as follows:
In Sec.~\ref{SKT}, 
we introduce a quantum kicked top and its classical 
counterpart.
Although this system is well-known in the literature,
we again mention them for this paper to 
be self-contained. 
In Sec.~\ref{CKTs}, we introduce the 
coupled kicked tops (CKTs).
We also introduce the von Neumann and linear 
entropies of the subsystem as measures 
of entanglement. 
We numerically find that, 
for {\it both} von Neumann and linear entropies, 
the time variation is nearly a linear function
of time when the nonlinearity parameter $k$ is large enough.
We also find that, 
when the coupling is weak, the production rate of 
the linear entropy is nearly 
proportional to $\epsilon^2$ where $\epsilon$ is the 
interaction strength between two tops.
This result implies that a perturbative treatment is possible. 
In Sec.~\ref{sec:perturb}, 
we derive a perturbative expression 
for the linear entropy of CKTs.
It consists of a  
correlation function for a single kicked top
which decays rapidly when the kicked top is chaotic. 
We also compare numerical results with the perturbative 
expression, and show that the agreement is good 
up to the long time where the entanglement production 
rate can be defined.
In Sec.~\ref{sec:discussion}, using the formula in the previous section, 
we show that the increment of the strength of chaos does not enhance 
the entanglement production rate in the strongly chaotic region. 
We also numerically confirm that this phenomenon actually 
happens for CKTs~\cite{TFM02}.  
The relationship between our result 
and previous results for the weakly chaotic region is discussed 
in Sec.~\ref{sec:comparison}.
Finally, we summarize this paper in Sec.~\ref{sec:summary}.

\section{Quantum and classical kicked top}
\label{SKT}

A kicked top~\cite{Haake00} is described by the following Hamiltonian:
\begin{equation}
\label{eq:defTopH}
H(t)= \frac{k}{2j} J_z^2 \sum_n \delta(t-n) + \frac{\pi}{2} J_y, 
\end{equation}
where $(J_x, J_y, J_z)$ are angular momentum operators
that satisfy $[J_x,J_y] = i J_z$ etc., $j$ is their magnitude, 
which is a conserved quantity, and $k$ is the nonlinear parameter.
Since we take $\hbar=1$, $1/j$ plays an effective Planck's constant.
The nonlinear parameter $k$ changes the qualitative nature of the
classical dynamics (see Eqs.~(\ref{eq:cldyn}) below): On one hand, the
classical kicked top exhibits chaotic behavior, i.e. the phase 
space of the classical top is dominated by chaotic seas, when $k
\gtrsim 3$. On the other hand, the classical phase space is dominated
by tori when $k \lesssim 2.5$~\cite{Haake00}.
A Floquet operator (i.e. a one-step time-evolution operator)
corresponding to the Hamiltonian~(\ref{eq:defTopH}) is  
\begin{equation}
\label{eq:defTopU}
U=\exp ( -i k J_z^2/j ) \exp (-i \pi J_y/2 ).
\end{equation}
In the numerical computations, we employ the
$|jm\rangle$-representation, where 
\mbox{$J^2 |jm \rangle = j(j+1)|jm \rangle$} and 
$J_z |jm \rangle = m |jm \rangle$.
The $|jm\rangle$-representation of $U$ (Eq.~(\ref{eq:defTopU})) 
is (note that $j$ is a conserved quantity), 
\begin{equation}
U_{m'm}
=\langle jm' | e^{-i \pi J_y/2} | jm \rangle
e^{-i k m^2/(2j)}
=d_{m'm}^{(j)}(\pi/2)
e^{-i k m^2/(2j)}
\end{equation}
where $d_{m'm}^{(j)}(\beta)$
is the Wigner rotational matrix~\cite{Sakurai94,VMK88}:
\begin{eqnarray}
d_{m'm}^{(j)}(\beta)
&=& \sum_l (-1)^{l-m+m'}
\frac{\sqrt{(j+m)!(j-m)!(j+m')!(j-m')!}}
{(j+m-l)!l!(j-l-m')!(l-m+m')!}
\nonumber
\\
&&\times
\cos^{(2j-2l+m-m')}(\beta/2)
\sin^{(2l-m+m')}(\beta/2).
\end{eqnarray}
In the numerical evaluation of the Wigner matrix, we employ
its Jacobi polynomial expression~\cite{VMK88}.

The corresponding classical dynamics is described by the following
mapping 
\begin{subequations}
\label{eq:cldyn}
\begin{eqnarray}
x' &=& z \cos(k x)+ y \sin(k x),\\
y' &=& -z \sin(k x)+ y \cos(k x),\\
z' &=& -x,
\end{eqnarray}
\end{subequations}
where $x=J_x/j$, $y=J_y/j$, and $z=J_z/j$.
Note that $x^2+y^2+z^2$ is a conserved quantity, since $j$ is conserved.
We hence employ polar coordinates 
$\theta= \cos^{-1} z$ and $\phi = \tan^{-1} (y/x)$ to concisely 
describe both classical and quantum dynamics.

In studying the correspondence between quantum and classical
dynamics, it is useful to employ a spin-coherent state
$|\theta,\phi \rangle$~\cite{spincoherent}
\begin{equation}
\langle j m| \theta,\phi\rangle
=\frac{\gamma^{j-m}}{(1+|\gamma|^2)^j} \sqrt{\frac{2j!}{(j+m)!(j-m)!}}
\end{equation}
where $\gamma=e^{i \phi} \tan (\theta/2)$.
We accordingly employ the Husimi representation of a state vector
$|\Psi \rangle$: 
\begin{equation}
Q(\theta,\phi)=|\langle \theta,\phi|\Psi \rangle|^2.
\end{equation}
To study the classical counterpart of the spin-coherent
state $|\theta_0,\phi_0\rangle$, we employ a ``Gaussian'' distribution
function on the classical phase space $(\theta,\phi)$
\begin{equation}
f_0(\theta,\phi) 
\sim 
e^{-(\theta-\theta_0)^2/2 \sigma^2 
-(\phi-\phi_0)^2/2 \sigma^2}
\label{dist2}
\end{equation}
where the fluctuation $\sigma=1/\sqrt{j}$ is determined by the Husimi
function of $|\theta_0, \phi_0 \rangle$.
\begin{figure}
\includegraphics[scale=1.0]{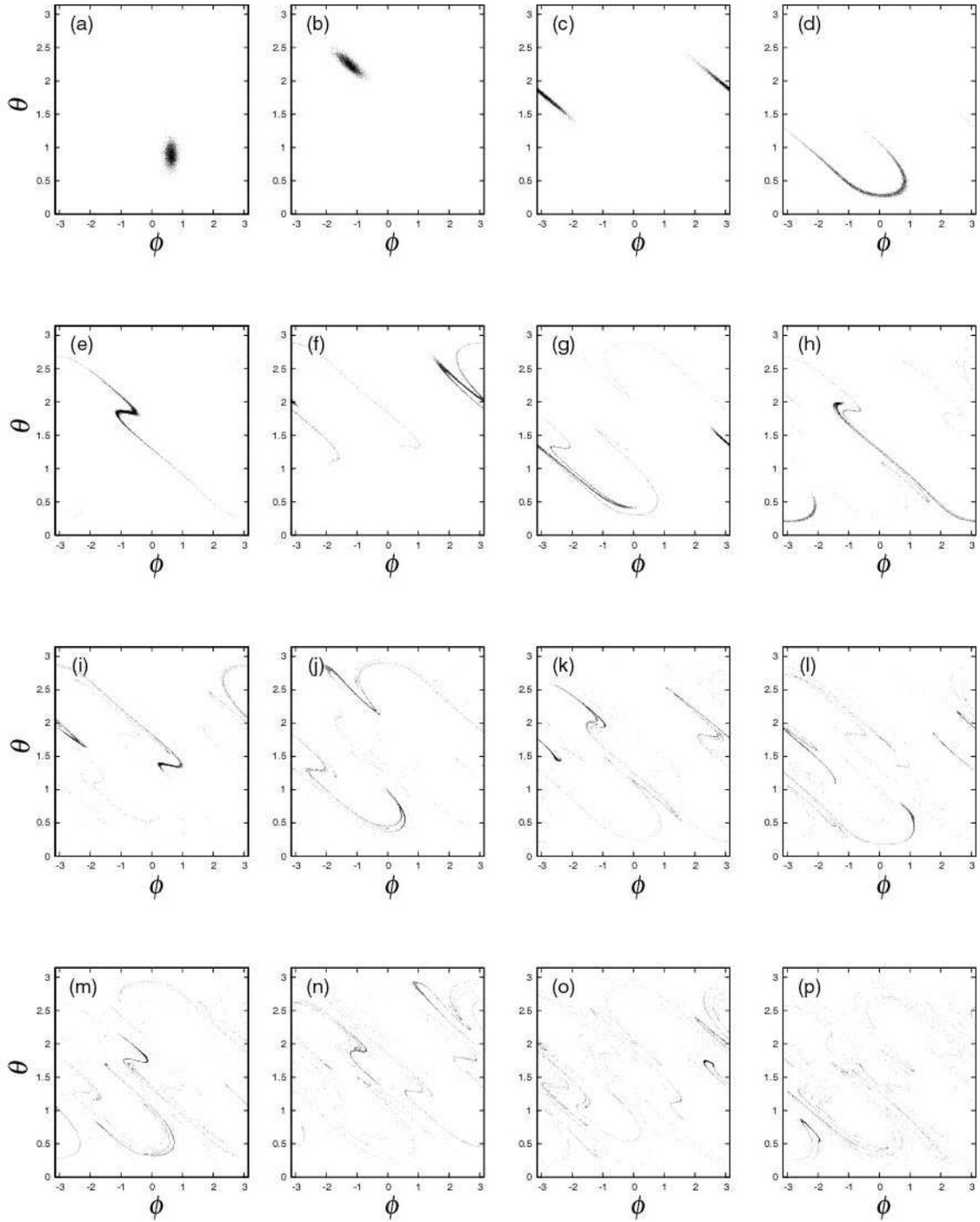}
\caption{\label{fig:top1d_cl}
Classical dynamics of the distribution function with $k=3.0$.
(a), (b), \ldots, (p) corresponds to 
$t=0, 1,\ldots, 15$, respectively.
The center of the initial ``Gaussian'' distribution function
(width $\sigma=0.1$) is located at $(\theta_0, \phi_0)=(0.89,0.63)$.
}
\end{figure}

\begin{figure}
\includegraphics[scale=1.0]{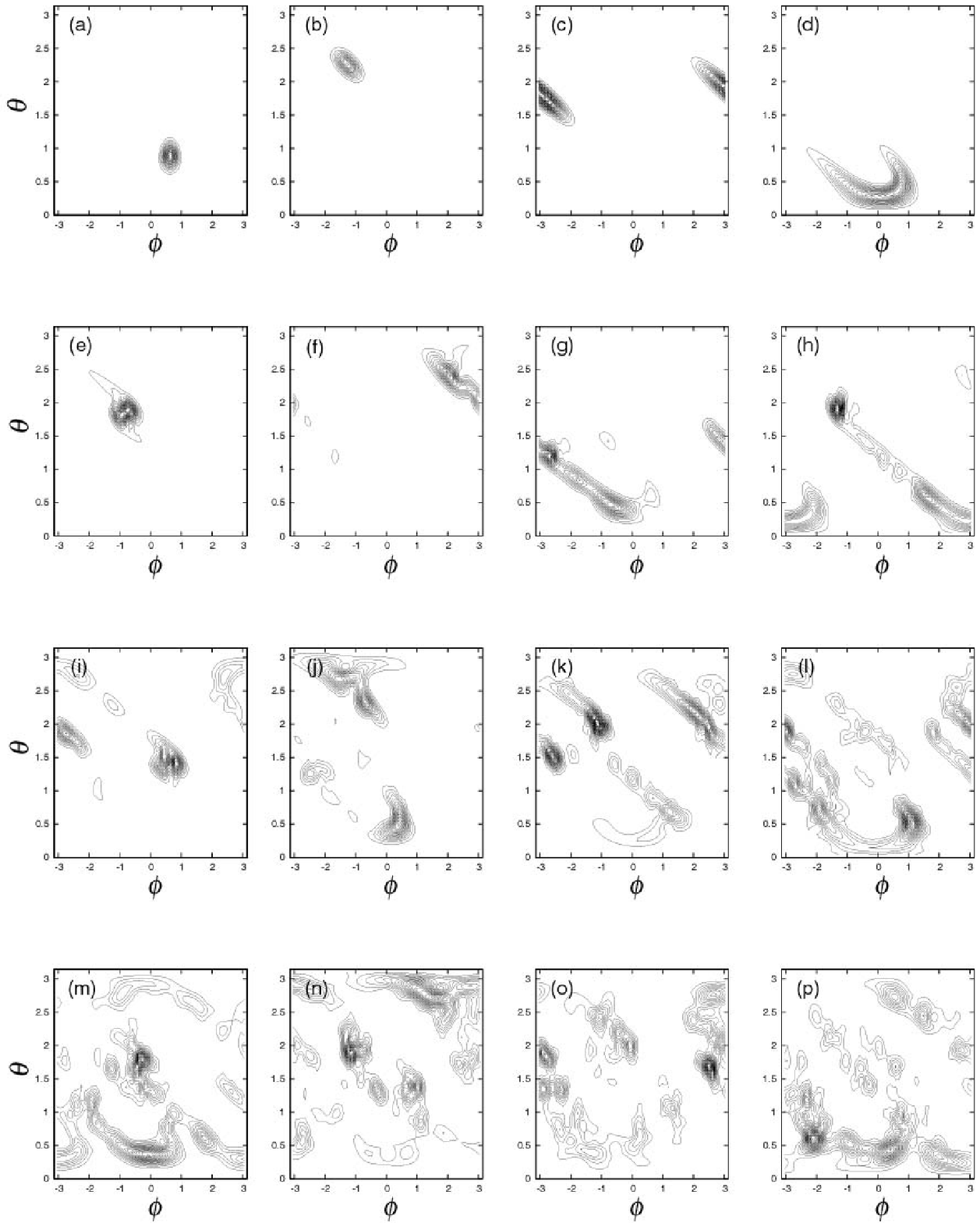}
\caption{\label{fig:top1d_qu}
Quantum dynamics of the Husimi function with $j=80$ and $k=3.0$.
(a), (b), \ldots, (p) corresponds to $t = 0, 1, \ldots, 15$,
respectively. 
The center of the initial spin-coherent state 
is located at $(\theta_0, \phi_0)=(0.89,0.63)$.
}
\end{figure}

\begin{figure}
\includegraphics[scale=0.9]{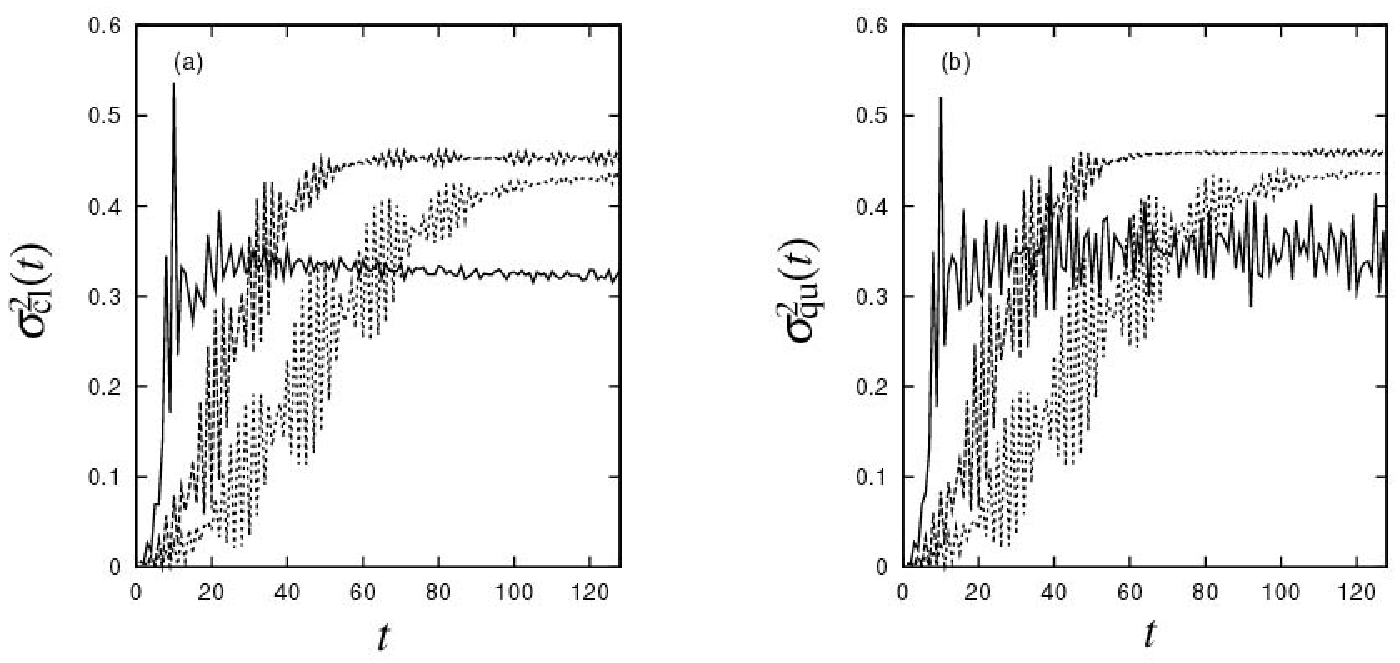}
\caption{\label{fig:momentum}
Time evolution of the variances for
(a) the classical and (b) quantum kicked tops.
The dotted, dashed, and solid lines correspond 
to $k=0.5$, $1.0$, and $3.0$, respectively.
The initial states of (a) and (b) are the same as in 
Figs.~\ref{fig:top1d_cl} and \ref{fig:top1d_qu}, 
respectively.
Since the regular evolutions ($k=0.5,1.0$) are trapped by tori 
whose sizes are large in the $J_z$ direction, 
the variances takes larger
value than that for the chaotic case ($k=3.0$).
}
\end{figure}

Figures \ref{fig:top1d_cl} and \ref{fig:top1d_qu}
show the classical dynamics of the distribution function
and quantum dynamics of the Husimi function for the kicked top
in a semiclassical regime ($j=80$), respectively.
As is mentioned above, at the initial stage of the dynamics, 
the quantum-classical correspondence holds well. That is, both quantum
and classical distribution functions behave very similarly.
Such a precise correspondence between the distribution
functions is lost as time elapses, due to quantum
interference.
However, even in a much longer time period, 
the variances of $z(t) = J_z(t)/j$, 
\begin{eqnarray}
\sigma_{\rm cl}^2(t)&=&
\langle z(t)^2 \rangle_{\rm cl} -\langle z(t) \rangle_{\rm cl}^2,
\\
\sigma_{\rm qu}^2(t)&=&
\langle z(t)^2 \rangle_{\rm qu} -\langle z(t) \rangle_{\rm qu}^2,
\end{eqnarray}
have a good quantum-classical
correspondence. See Fig.~\ref{fig:momentum}. 
Here $\langle\ldots\rangle_{\rm cl}$ and 
$\langle\ldots\rangle_{\rm qu}$ are the expectation values for
the classical and the quantum systems, respectively.
An explanation in terms of the classical phase-space dynamics is as
follows:
In the classically regular region ($k=0.5, 1.0$), 
trajectories are trapped by tori, and the variances
exhibits periodic modulations. According to the sizes (along the $J_z$
direction) of the trapping tori, the variances takes various
values. Furthermore, the variances exhibit recurrence phenomena within
a rather short time period (not shown here).  
On the other hand, in the classically chaotic region ($k=3.0$), both 
$\sigma_{\rm cl}^2(t)$ and $\sigma_{\rm qu}^2(t)$ increase 
rather quickly 
and saturate to the value $\sigma_{\rm sat}^2 = 1/3$ which 
is estimated by assuming a uniform distribution.
In other words, the phase-space
distribution functions spread uniformly all over the whole phase
space, which is bounded in the case of the top.

\section{Numerical experiment of quantum kicked tops}
\label{CKTs}


Here we return to our motivation: 
How does classical chaotic dynamics 
affect the entanglement production of the corresponding quantum system? 
To investigate this issue, we employ coupled kicked tops (CKTs),
which is introduced by Miller and Sarkar~\cite{MS99}, 
as a target system of numerical experiments.
The CKTs are described by the following Hamiltonian: 
\begin{eqnarray}
\label{eq:defCTH}
H(t) = H_1(t)+H_2(t)+H_{\epsilon}(t)
\end{eqnarray}
where 
\begin{eqnarray}
H_1(t) &=&
\frac{k_1}{2j} J_{z_1}^2 \sum_n \delta(t-n) 
+\frac{\pi}{2} J_{y_1},
\\
H_2(t) &=&
\frac{k_2}{2j} J_{z_2}^2 \sum_n \delta(t-n) 
+\frac{\pi}{2} J_{y_2}, 
\\
\label{eq:Hint}
H_{\epsilon}(t)
&=&
\frac{\epsilon}{j} J_{z_1} J_{z_2} 
\sum_n \delta(t-n),
\end{eqnarray}
with $[J_{x_l},J_{y_m}]=i J_{z_l} \delta_{lm}$ etc. ($l,m=1,2$).
Here $k_l$ is the nonlinear parameter of the $l$-th top,
and $\epsilon$ is the strength of the coupling between these two tops.
Corresponding to the Hamiltonian, Eq.~(\ref{eq:defCTH}), 
we employ a Floquet operator (a one-step time-evolution operator) 
\begin{equation}
  \label{eq:defCKTU}
  U \equiv U_{\epsilon} U_1 U_2
\end{equation}
where 
$U_1 =  e^{-i k_1 J_{z_1}^2/2j} e^{-i \pi J_{y_1}/2}$,
$U_2 = e^{-i k_2 J_{z_2}^2/2j} e^{-i \pi J_{y_2}/2}$
and 
$U_{\epsilon} = e^{-i \epsilon J_{z_1} J_{z_2}/j}$.

Since we will consider only the case where 
the density operator of the total system at time $t$, 
$\rho(t)$, describes a pure state, 
we employ entropies of subsystems as measures of quantum
entanglement~\cite{EntanglementMeasure}. 
More precisely, we employ von
Neumann and linear entropies of the first top:
\begin{eqnarray}
S_{\rm vN}(t) &=& -{\rm Tr}_1 \{ \rho^{(1)}(t) \ln \rho^{(1)}(t) \},
\\
\label{eq:defSlin}
S_{\rm lin}(t) &=& 1- {\rm Tr}_1 \{ \rho^{(1)}(t)^2 \},
\end{eqnarray}
where $\rho^{(1)}(t)={\rm Tr}_2 \{ \rho(t) \}$ is the reduced density
operator for the first top.
Note that the von Neumann entropy (or the linear entropy) of the
second top takes the same value as that of the first top, when the
whole system is in a pure state. 
To calculate $S_{\rm vN}(t)$ and $S_{\rm lin}(t)$ numerically, we use
the eigenvalues $\lambda_i(t)$ of $\rho^{(1)}(t)$ as
\begin{eqnarray}
S_{\rm vN}(t) &=& -\sum_i \lambda_i(t) \ln \lambda_i(t),
\\
S_{\rm lin}(t) &=& 1- \sum_i \{\lambda_i(t)\}^2.
\end{eqnarray}

We numerically examine the productions of quantum entanglement,
using separable states as initial states. In particular,
we focus on the the system parameter dependence, i.e. $k_1$, $k_2$
and $\epsilon$ dependence, of the entanglement productions 
(measured by the entropies of the subsystems).
For simplicity, we only show the cases where $k=k_1=k_2$.
As an initial state, we employ the following pure and separable state 
\begin{equation}
\label{eq:initial}
\rho(0)= | \Psi(0) \rangle \langle \Psi(0)|,
\end{equation}
where  
$| \Psi(0) \rangle =
|\theta_1, \phi_1 \rangle \otimes|\theta_2, \phi_2 \rangle$
is a direct product of spin coherent states.
%
In studying the chaotic region, the center of the initial
spin coherent state is located in the chaotic sea. 
Even such numerical experiments with a restricted class of initial
states provides an insight about the {\em typical} behavior of
chaotic CKTs, when the fraction of tori is small in
phase space for the corresponding classical system 
(in the case of CKTs, $k>3.0$).

\begin{figure}
\includegraphics[scale=0.9]{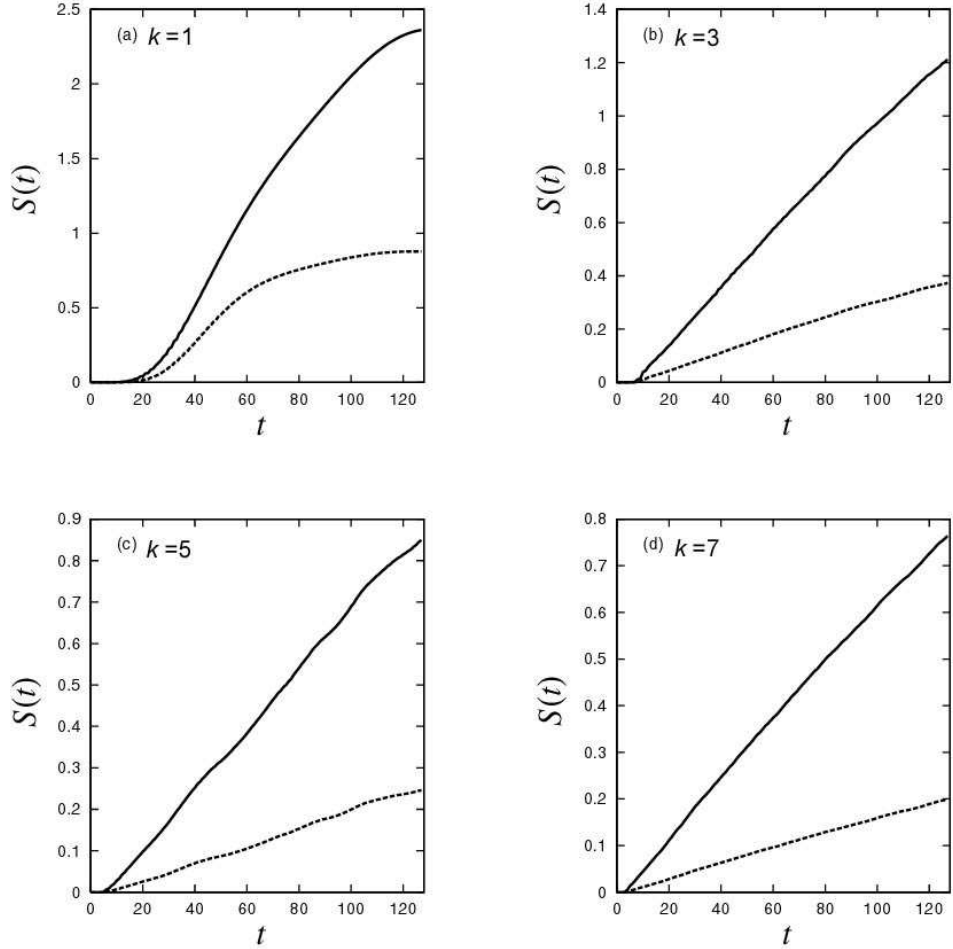}
\caption{\label{fig:k-dep}
Time evolutions of the linear (dashed line) 
and von Neumann (solid line) entropies.
Parameters are $\epsilon=10^{-3}$ (in an weak coupling region)
and $j=80$ (in a semiclassical region).
The initial state is Eq.~(\ref{eq:initial})
with $\theta_1=\theta_2=0.89$ and $\phi_1=\phi_2=0.63$.
The center of the initial state is located 
in the chaotic seas for the cases of (b), (c), and (d).
}
\end{figure}

Figure \ref{fig:k-dep} shows the time evolutions of the entropies
with various values of $k$. 
The entropies stick to nearly zero 
until the ``rising time'', 
and then they increase nearly linearly as 
a function of time for chaotic cases 
(see Figs.~\ref{fig:k-dep} (b),(c),(d)). 
Though not shown here, 
they saturate to finite values after the long time evolution, due to
the finiteness of the dimension of the Hilbert space.
In the following, we focus on the ``intermediate'' region 
where the entropies increase monotonically as a function of time.
Our extensive numerical experiments conclude that, in the
intermediate region, the linear entropy as well as the von Neumann
entropy increases nearly {\it linearly} as a function of time, 
when the chaos is strong enough and the coupling is weak enough.

\begin{figure}
\includegraphics[scale=0.9]{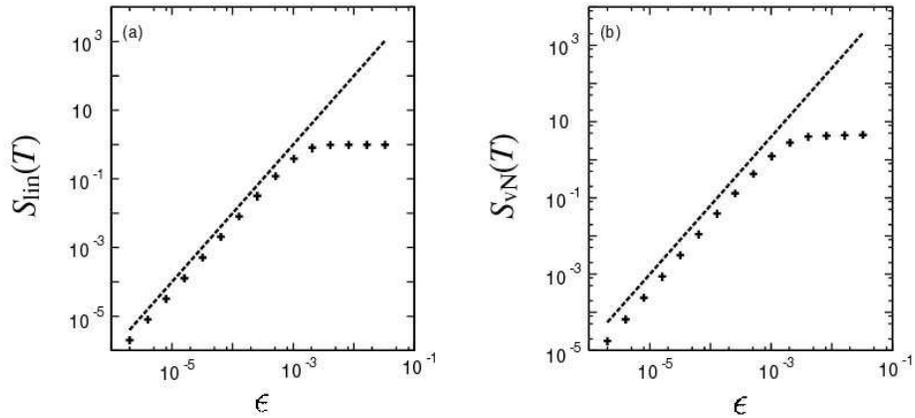}
\caption{\label{fig:eps-dep}
$\epsilon$ dependence of (a) the linear and 
(b) von Neumann entropies at $T=128$ with $k=3.0$.
The dashed lines denote $10^6 \epsilon^2$ and 
$10^6 \epsilon^{1.8}$ for (a) and (b), respectively.
The initial state is the same as in Fig.~\ref{fig:k-dep}.
}
\end{figure}

Figure~\ref{fig:eps-dep} 
shows $\epsilon$ dependence 
of the linear and von Neumann entropies at $T=128$.
When the coupling is weak (i.e., $\epsilon < 10^{-3}$), the linear  
entropy obeys a perturbative behavior 
$S_{\rm lin}(T)\propto \epsilon^2$ (see Fig.~\ref{fig:eps-dep}(a)).
At the same time, the time step $T=128$ belong to the region where 
the linear entropy increases linearly in time.
In contrast to this, when $\epsilon$ is much larger than $10^{-3}$, 
the entropy does not belong to the perturbative region,
and saturates to a finite value, which is determined by the finite
size of the Hilbert space of CTKs.
These observations suggests us to analyze 
the behavior of the linear entropy 
using a perturbation treatment for the 
interaction strength $\epsilon$.
This is the subject of the next section.
On the other hand, 
we confirmed that 
$S_{\rm vN}(T) \propto \epsilon^{1.8}$ 
when $\epsilon$ is small enough 
(Fig.~\ref{fig:eps-dep} (b)).
It seems that an usual perturbative treatment is difficult
to explain the exponent $1.8$, so we will concentrate 
on the linear entropy below.

\section{Perturbative expression for the linear entropy}
\label{sec:perturb}

\subsection{Perturbation treatment}

We evaluate $S_{\rm lin}(t)$, Eq.~(\ref{eq:defSlin}), 
the linear entropy of
the first top, by using the time-dependent perturbation theory with a
small parameter $\epsilon$. 
First, we introduce the interaction pictures, 
of the density matrix 
$\tilde{\rho}(t) = U_0^{\dagger}(t) \rho(t) U_0(t)$,
and, of the operator $\hat{A}(t)= U_0^{\dagger}(t) A U_0(t)$,
where $U_0(t)=(U_1 \otimes U_2)^t$. 
That is, $\hat{A}(t)$ corresponds to the ``free'' evolution of the
operator $A$.
Accordingly, the time evolution of $\tilde{\rho}(t)$ is described by
the unitary mapping
$\tilde{\rho}(t)=
\hat{U}_{\epsilon}(t)\tilde{\rho}(t-1) \hat{U}^{\dagger}_{\epsilon}(t)$,
where the expansion of $\hat{U}_{\epsilon}(t)$ by small $\epsilon$
takes the following form
\begin{equation}
\hat{U}_{\epsilon}(t) 
=1 - \frac{i \epsilon}{j} \hat{V}(t) 
-\frac{\epsilon^2}{2 j^2}
\hat{V}^2(t)
+{\cal O}(\epsilon^3)
\end{equation}
with $\hat{V}(t)=\hat{J}_{z1}(t) \hat{J}_{z2}(t)$.
Hence, the unitary mapping of $\tilde{\rho}(t)$ becomes
\begin{eqnarray}
\tilde{\rho}(t)
&=& \tilde{\rho}(t-1)
+\frac{i \epsilon}{j} [\tilde{\rho}(t-1), \hat{V}(t)]
-
\frac{\epsilon^2}{2 j^2}
[[\tilde{\rho}(t-1), \hat{V}(t)], \hat{V}(t)]
+{\cal O}(\epsilon^3).
\end{eqnarray}
By induction, we have
\begin{eqnarray}
\tilde{\rho}(t)
&=& {\rho}(0)
+\frac{i \epsilon}{j} 
\sum_{l=1}^{t} [{\rho}(0), \hat{V}(l)]
\nonumber
\\
&&
-\frac{\epsilon^2}{j^2}
\sum_{l=2}^{t} \sum_{m=1}^{l-1} [[ {\rho}(0), \hat{V}(l)],\hat{V}(m)]
-\frac{\epsilon^2}{2 j^2} 
\sum_{l=1}^{t} [[ {\rho}(0), \hat{V}(l)],\hat{V}(l)]
+{\cal O}(\epsilon^3).
\end{eqnarray}
By tracing out the second system, we have
\begin{eqnarray} 
\tilde{\rho}^{(1)}(t)
&=& {\rho}^{(1)}(0)
+\frac{i \epsilon}{j} 
\sum_{l=1}^{t} [{\rho}^{(1)}(0), \hat{J}_{z1}(l) ]
\langle \hat{J}_{z2}(l) \rangle_2
\nonumber
\\
&&
+\frac{\epsilon^2}{2 j^2}
\sum_{l=1}^{t} 
\left \{ 
[\hat{J}_{z1}(l){\rho}^{(1)}(0), \hat{J}_{z1}(l)] 
+
[\hat{J}_{z1}(l), {\rho}^{(1)}(0) \hat{J}_{z1}(l)]
\right \} 
\langle \hat{J}^2_{z1}(l) \rangle_2
\nonumber
\\
&&
+\frac{\epsilon^2}{j^2}
\sum_{l=2}^{t} \sum_{m=1}^{l-1}
\left \{
[ \hat{J}_{z1}(l), {\rho}^{(1)}(0) \hat{J}_{z1}(m) ] 
\langle \hat{J}_{z2}(m) \hat{J}_{z2}(l) \rangle_2
\right.
\nonumber
\\
&&
\left.
+
[ \hat{J}_{z1}(m) {\rho}^{(1)}(0), \hat{J}_{z1}(l) ]
\langle
\hat{J}_{z2}(l)
\hat{J}_{z2}(m) 
\rangle_2
\right \}
+{\cal O}(\epsilon^3)
\end{eqnarray}
where $\langle A \rangle_2 
= {\rm Tr}_2 \{ \rho^{(2)}(0) A \}$ 
is an average for subsystem 2. 
%
Finally, we obtain a second order perturbation formula of
$S_{\rm lin}(t) = 
S^{\rm PT}_{\rm lin}(t) + {\cal O}(\epsilon^3)$:
\begin{equation}
\label{eq:miyaji}
S^{\rm PT}_{\rm lin}(t) 
= S_0 \sum_{l=1}^t \sum_{m=1}^t D(l,m)
\end{equation}
where $S_0=2 \epsilon^2 j^2$ and $D(l,m)$ is a correlation function
of the uncoupled system. Since the interaction Hamiltonian 
$H_{\epsilon}(t)$, Eq.~(\ref{eq:Hint}), is in a bilinear form, $D(l,m)$ is
decomposed into a product of correlation functions of uncoupled subsystems
\begin{equation}
  \label{eq:defD}
  D(l,m) = C_1(l,m) C_2(l,m)
\end{equation}
where
\begin{equation}
\label{eq:defCi}
C_i(l,m) = \langle \hat{z_i}(l) \hat{z_i}(m) \rangle_i
- \langle \hat{z_i}(l)\rangle \langle \hat{z_i}(m) \rangle_i
\quad 
\end{equation}
and $\hat{z_i}(l)= \hat{J}_{z_i}(l)/j$ ($i=1,2$).
In the perturbation formula, Eq.~(\ref{eq:miyaji}), 
$S_0$ is a rather trivial factor implying
$S^{\rm PT}_{\rm lin}(t)\propto\epsilon^2$ as is observed in
Fig.~\ref{fig:eps-dep}(a). 
The nature of the dynamics for the tops is reflected in $D(l,m)$.

Let us remark important points of our perturbation formula:
(i) In common with the exact case, $S^{\rm PT}_{\rm lin}(t)$ has a
symmetric form about the exchange of the first and the second tops. 
That is, our perturbative treatment preserves this symmetry, although
we start from a perturbative treatment of the linear entropy of the
first top.
(ii) Our formula has a similarity with those in phenomenological
descriptions of linear irreversible processes~\cite{Kubo1985}, in the
sense that these theories use time correlation functions to describe
relaxation phenomena. This is useful both for making
phenomenological arguments and for establishing a link between 
a phenomenological theory and a microscopic theory
(cf. the linear response theory of nonequilibrium statistical
mechanics~\cite{Kubo1985});
(iii) Since our approach does not take into account the effect of the
recurrence, the formula~(\ref{eq:miyaji}) would have qualitatively
different applicability to the classically regular and chaotic
systems. For classically regular systems, our theory
would break down in relatively short time period, due to the smallness
of the period of the recurrence. On the other hand, for chaotic
systems, we numerically confirmed that our theory works 
for a rather long time period.

\subsection{Comparison with numerical results}

\begin{figure}
\includegraphics[scale=0.9]{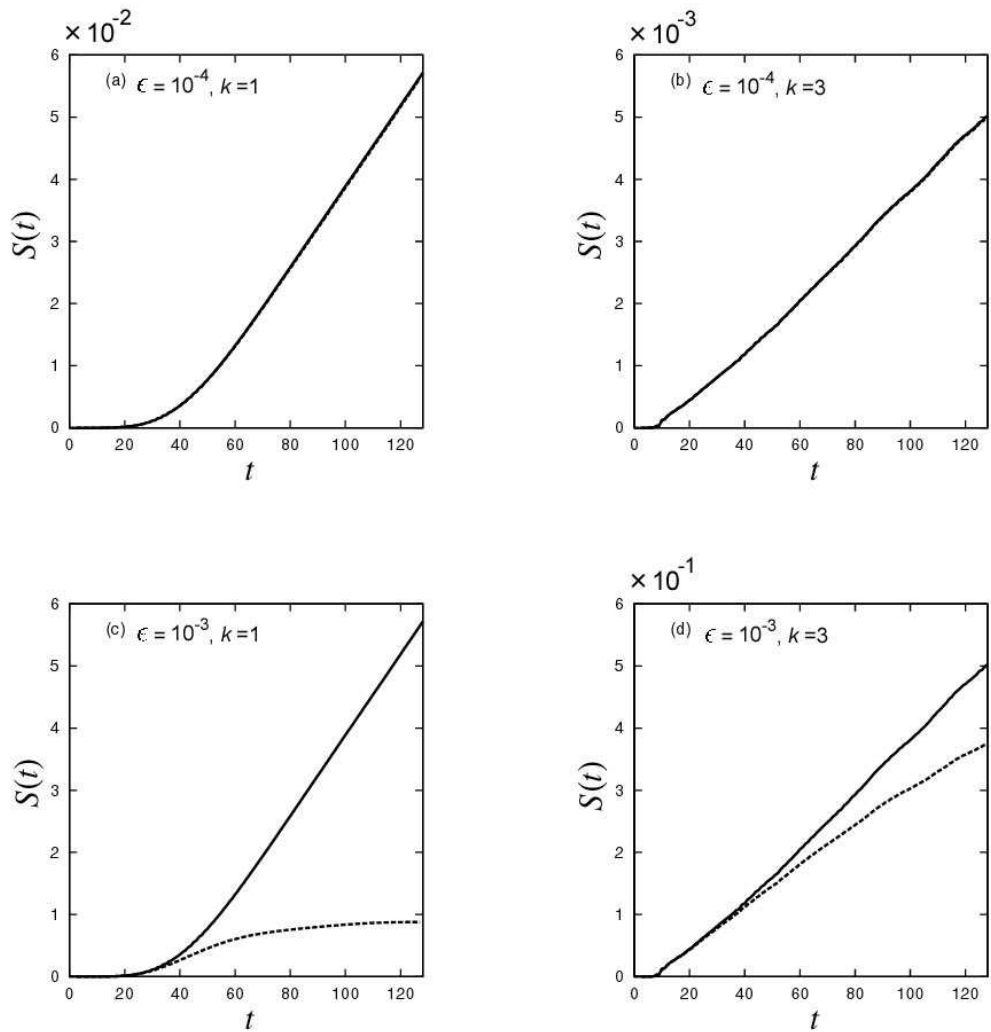}
\caption{\label{fig:miyaji}
Estimations of linear entropies by the perturbative
formula, Eq.~(\ref{eq:miyaji}). 
It is compared with the exact numerical result (in dashed lines).
The parameters are 
(a) $\epsilon=10^{-4}$ (perturbative region) with $k=1$ (regular)
(b) $\epsilon=10^{-4}$ (perturbative region) with $k=3$ (chaotic),
(c) $\epsilon=10^{-3}$ (intermediate region) with $k=1$ (regular),
(d) $\epsilon=10^{-3}$ (intermediate region) with $k=3$ (chaotic).
}
\end{figure}

We numerically examine our formula, Eq.~(\ref{eq:miyaji}).
In Fig.~\ref{fig:miyaji}, we plot both 
$S_{\rm lin}(t)$ and $S^{\rm PT}_{\rm lin}(t)$
for the intermediate coupling 
and weak coupling cases with regular and chaotic conditions.
The initial state is the direct product of 
the spin-coherent state as before. 
As shown in Fig.~\ref{fig:eps-dep}, 
$\epsilon=10^{-4}$ is the perturbative region, so 
the agreement between $S_{\rm lin}(t)$ and $S^{\rm PT}_{\rm lin}(t)$
is very good for different $k$'s up to $t \simeq 100$
(Figs.~\ref{fig:miyaji} (a) and (b)). 
Note that our perturbative expression works 
for such a long time to reproduce the linear increment of the
entanglement productions in time.
Such a correspondence degrades as $\epsilon$ gets 
larger, of course,  
as shown in Figs.~\ref{fig:miyaji} (c) and (d).  
However, as far as concerning the chaotic case $k=3.0$,
our expression describes the entanglement production,
at least, qualitatively.

\section{Dynamical aspects of entanglement}
\label{sec:discussion}

\subsection{Harder chaos does not mean larger entanglement production}
\label{sec:harderChaos}

In this section,
the perturbative formula, Eq.~(\ref{eq:miyaji}), 
is employed to answer the following
question: how does the strength of chaos influence on 
the entanglement production rate in 
the strongly chaotic regions where the influences
from tori are negligible.

We examine $C_i(l,m)$, Eq.~(\ref{eq:defCi}), which describes the
fluctuation of $z_i = J_{z_i}/j$. Since the tops are strongly chaotic, 
we impose several phenomenological assumptions on $C_i(l,m)$.
Since the phase space of the kicked top is bounded, the distribution
function in the phase space becomes quickly uniform in the strongly
chaotic region. Hence we assume 
$C_i(l,l) \simeq \sigma_{\rm sat}^2$,
where we ignore a short transient 
before the distribution function becomes uniform (see
Fig.~\ref{fig:momentum}). 
The magnitude of the fluctuation $\sigma_{\rm sat}^2=1/3$
is determined by the assumption that the distribution function 
is uniform on sphere $(\theta,\phi)$.
The boundedness of the phase space allows us to employ another
assumption that the relaxation of $C_i(l,m)$ ($l\ne m$) is exponential
with an exponent $\gamma_i$ \cite{BK83}.
Furthermore, it is natural to assume that the exponent $\gamma_i$
becomes larger as the positive Lyapunov exponent of the
corresponding classical system becomes larger. 
The simplest function that satisfies the assumptions above is
\begin{equation}
\label{eq:correlation}
C_i(l,m) \simeq \sigma_{\rm sat}^2 e^{-\gamma_i|l-m|}. 
\end{equation}
Hence $D(l,m)$, Eq.~(\ref{eq:defD}), becomes
\begin{equation}
  \label{eq:Dpheno}
  D(l,m) \simeq  D_0 e^{-\gamma|l-m|} 
\end{equation}
where $D_0 = \sigma_{\rm sat}^4$ and $\gamma=\gamma_1+\gamma_2$.
Accordingly, Eq.~(\ref{eq:miyaji}) provides the following evaluation
of the linear entropy 
\begin{equation}
S^{\rm PT}_{\rm lin}(t) 
\simeq S_0 D_0 
\left[
\coth (\gamma/2) t
-\frac{1-e^{-\gamma t}}{\sinh \gamma -1}
\right].
\end{equation}
When the relaxation time of $D(l,m)$ is much shorter than the time scale of
the stationary entanglement production region, we have an
{\em entanglement production rate} $\Gamma$
\begin{equation}
\label{eq:dMdt}
\Gamma \equiv
\left. \frac{d S^{\rm PT}_{\rm lin}(t)}{dt} \right|_{t \gg 1/\gamma} 
\simeq
\Gamma_0 \coth (\gamma/2),
\end{equation}
where $\Gamma_0 \equiv S_0 D_0$.
From this relation, it is shown that 
$\Gamma$ decreases as $\gamma$ becomes larger, i.e. the chaos of the
corresponding classical system becomes stronger. 
That is, {\it the increment of the strength of chaos does not enhance
the production rate of entanglement}. 
Furthermore, in the limit $\gamma \rightarrow \infty$, 
$\Gamma$ quickly {\it saturates} to a finite value, $\Gamma_0$.

At first glance, our prediction seem to be counter-intuitive.
Hence we provide an explanation of the prediction to summarize this
subsection: 
The entanglement productions are induced by the fluctuation
of the interaction Hamiltonian $H_{\epsilon}(t)$,
in the interaction picture.
Since the time dependence of $H_{\epsilon}(t)$ looks like very
``random'' in classically chaotic systems,
the contribution from $H_{\epsilon}(t)$ to the linear entropy
is reduced due to {\it dynamical averaging}.
We note that this mechanism is similar to that of the so called 
{\it motional narrowing} in spin relaxation
phenomena~\cite{Kubo1985,Slichter}.

\subsection{Properties of the correlation function 
and linear entropy production rate --- 
Saturation of the entanglement production}

\begin{figure}
\includegraphics[scale=0.9]{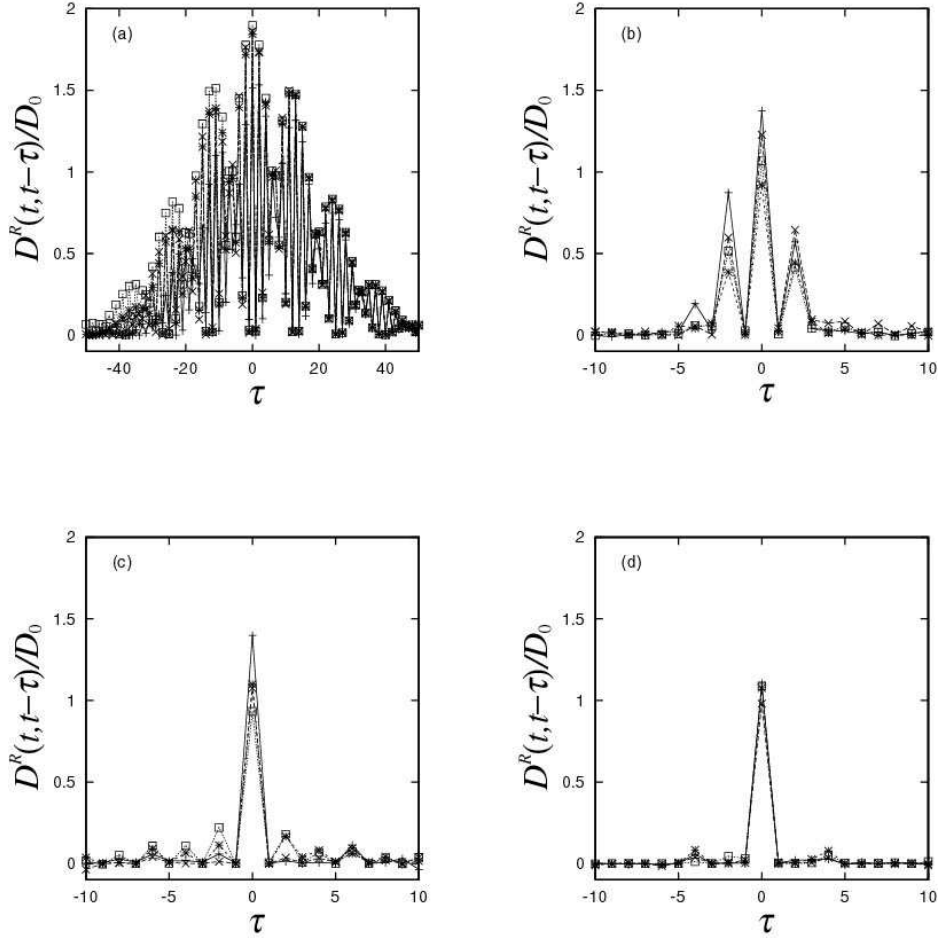}
\caption{\label{fig:correlation1}
Plot of $D^R(t,t-\tau)/D_0$ ($\equiv \Re D(t,t-\tau)/D_0$)
as a function of $\tau$ with different $t$'s ($t=40,50,60,70$)
and $k$'s. 
(a), (b), (c), (d) corresponds to $k=1$, $3$, $5$, $7$, respectively.
The initial state is Eq.~(\ref{eq:initial}) 
with $\theta_1=\theta_2=0.89$ and $\phi_1=\phi_2=0.63$. 
}
\end{figure}

We test the prediction of the phenomenological argument above
with numerical experiments.
First, we examine the correlation function $D$, Eq.~(\ref{eq:defD})
(see Fig.~\ref{fig:correlation1}).
We confirmed the assumption, Eq.~(\ref{eq:Dpheno}), for $D$
when the classical counterpart is chaotic ($k\ge3.0$):
The correlation function decays very quickly, althogh
it is difficult to detemine the exponent $\gamma$ directly from the
numerical evaluation of $D$ (see another estimation of $\gamma$ below).
On the other hand, the value of $D(t,t)$ is almost
independent with $k$, and approximately equal to 
$D_0=\sigma_{\rm sat}^4 = 1/9$.

\begin{figure}
\includegraphics[scale=0.9]{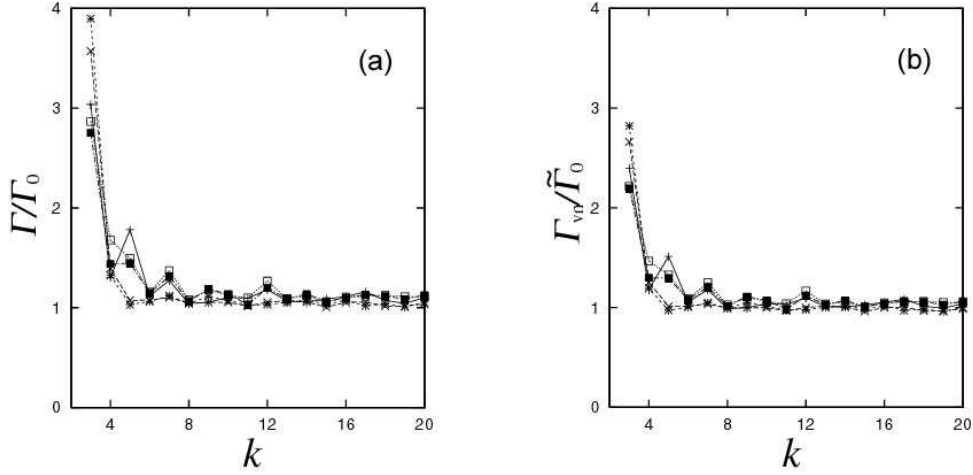}
\caption{\label{fig:k-dep2}
(a) $k$ dependence of the normalized linear entropy production 
rate 
for various initial conditions located in the chaotic sea.
Note that $\Gamma$ saturates to $\Gamma_0$ as $k$ increases.
(b) $k$ dependence of the normalized von Neumann 
entropy production rate $\Gamma_{\rm vN}$
for various initial conditions (the same as those of (a)) 
located in the chaotic sea.
Note that $\Gamma_{\rm vN}$ is scaled by 
$\tilde{\Gamma}_0 \equiv 2 \epsilon^{1.8} j^2 D_0$
instead of $\Gamma_0 = 2 \epsilon^{2} j^2 D_0$
(cf. Fig.~\ref{fig:eps-dep}).
}
\end{figure}

Second, we examine the nonlinear parameter $k$-dependence of the
entanglement production rate $\Gamma$, Eq.~(\ref{eq:dMdt}).
For each initial states, whose center of the spin-coherent
states is placed in the chaotic sea, we obtain $\Gamma$ using the least
square fitting for the time region from $t=20$ to 100,
where $t$-linearity holds (Fig.~\ref{fig:k-dep2}).
Hence we confirm that the increment of the strength of chaos 
does not enhance entanglement production rate 
in the perturbative regime, where $\epsilon$ is small enough.
It is also confirmed that the entropy production rate $\Gamma$
saturates to $\Gamma_0$ for large $k$,
which is also consistent with Eq.~(\ref{eq:dMdt}).
At the same time, we numerically find that $\Gamma_{\rm vn}$, the
entanglement production rate measured by von Neumann entropy
also exhibits a saturation in the large $k$ limit. Although we do not
have any analytical theory for  $\Gamma_{\rm vn}$, we expect that the
saturation of $\Gamma_{\rm vn}$ is explained by the similar explanation 
as that of the linear entropy $\Gamma$ (see Sec.~\ref{sec:harderChaos}).

\begin{figure}
\includegraphics[scale=1.0]{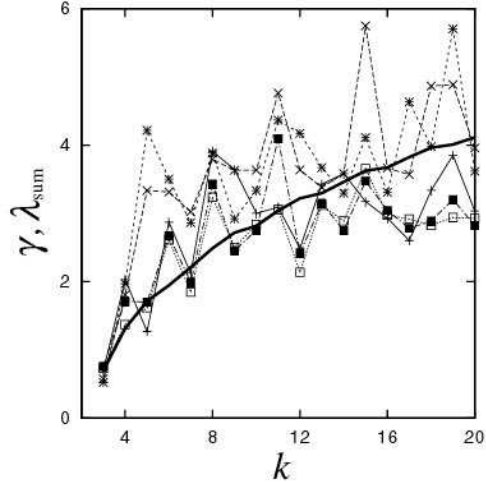}
\caption{\label{fig:k-dep3}
$k$ dependence of the correlation decay rate $\gamma$,
which is determined from Eq.~(\ref{eq:gammaeff}),
for various initial conditions located 
in the chaotic sea (scattered symbols).
The sum of the Lyapunov exponent $\lambda_{\rm sum}$
is also shown as a bold curve, which correspond to a single 
initial condition. 
We note that $\lambda_{\rm sum}$ depends on the initial 
condition only very weakly.
}
\end{figure}

In Fig.~\ref{fig:k-dep3}, we plot $k$-dependences of two quantities:
One is $\lambda_{\rm sum}\equiv \lambda_1 + \lambda_2$, where
$\lambda_i$ is the short-time (up to $t=100$) and 
phase-space averaged Lyapunov exponent  
for the initial classical distribution of $i$-th subsystem.
%
%
We note that $\lambda_{\rm sum} = 2 \lambda_1$ since $k_1 = k_2$.
The other is the decay rate $\gamma$ of the correlation function
$D$, Eq.~(\ref{eq:defD}). We estimate $\gamma$ from the phenomenological
estimation of $\Gamma$, Eq.~(\ref{eq:dMdt}), i.e.
\begin{equation}
  \label{eq:gammaeff}
  \gamma = 
  \ln \left( \frac{\Gamma/\Gamma_0+1}{\Gamma/\Gamma_0-1} \right),
\end{equation}
instead of the direct estimation from the assumption, 
Eq.~(\ref{eq:Dpheno}).
Figure \ref{fig:k-dep3} suggests $\gamma \simeq \lambda_{\rm sum}$.
This shows an evidence that the decay rates of the correlations of
tops are determined by the positive Lyapunov exponents of the
classical counterparts.
Thus our numerical experiments confirm the estimation, 
Eq.~(\ref{eq:dMdt}),
and it is concluded that the entanglement production rate is not increased 
by the increment of the strength of chaos in the strongly chaotic
region, and saturate to a finite value in the strong chaos limit.

Finally, we point out that it is natural to generalize our study on CKTs
to any strongly chaotic system with bounded phase space.
We note that we have already confirmed this for coupled kicked
rotors~\cite{SKO96,Lakshminarayan01,TAI89} 
with periodic boundary conditions of both position and momentum
coordinates.

\subsection{An extension to flow systems}
\label{sec:flow}

In this subsection,
we extend the above argument to {\it flow systems}
with continuous time.
Consider the case where flow systems are weakly coupled.
When the initial state is a pure product state (as is the case above),
the linear entropy produced in the composite system is
\begin{equation}
S^{\rm PT}_{\rm lin}(t) \simeq 
S_0 \int_0^t d \tau \int_0^t d \tau' D(\tau,\tau'),
\label{eq:flow}
\end{equation}
where $D(\tau,\tau')$ is a correlation function determined by the 
form of the interaction Hamiltonian.
When the subsystems are strongly chaotic with bounded phase space, 
we assume again
\begin{equation}
D(\tau,\tau') \simeq  D_0 e^{-\gamma|\tau-\tau'|}. 
\end{equation}
Substituting this into Eq.~(\ref{eq:flow}), we have
\begin{equation}
\label{eq:estGammaFlow}
\Gamma \simeq 
\frac{2 S_0 D_0}{\gamma}(1-e^{-\gamma t})
\rightarrow \frac{2 S_0 D_0}{\gamma} \quad (t \gg 1/\gamma).
\end{equation}
Hence, if the subsystem is strongly chaotic 
(i.e. $\gamma \rightarrow \infty$), the entanglement production rate
becomes zero. 
That is, \emph{strong chaos completely suppresses entanglement 
production!}
Although we have not numerically confirmed this suppression yet,
the similar suppressions of quantum relaxations in strongly chaotic
systems have been observed by Prosen and
\v{Z}nidari\v{c}~\cite{Prosen:PRE-65-036208,Znidaric:qph-0209145}. 
We will discuss this point further in Sec.~\ref{sec:summary}.

\section{Discussion on a weakly chaotic region}
\label{sec:comparison}

\begin{figure}
\includegraphics[scale=0.9]{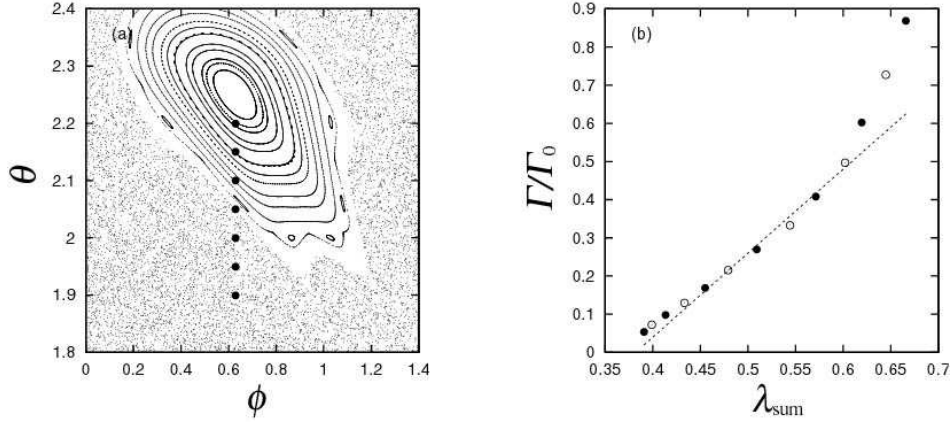}
\caption{\label{fig:LsVsGweak}
(a) A magnification of Poincare section of the kicked top with
$k=3.0$. 
(b) The correlation between $\lambda_{\rm sum}$ and $\Gamma$ in a
weakly chaotic region. 
In particular, solid circles correspond to those in (a).
A linear fitting is depicted by a dashed line
($\Gamma/\Gamma_0 = a \lambda_{\rm sum}+b$ with $a=2.2$ and $b=-0.84$).
Parameters are $\epsilon=10^{-4}$ (perturbative region) and $j=80$.
The center of the initial state of the first top 
is located at $(\phi_1,\theta_1)= (0.63,0.89)$, and 
those of the second top are depicted by solid circles in (a).
}
\end{figure}

In the previous section, we investigated the strongly chaotic region
of the CKTs and concluded that the increment of the strength of chaos
does not enhance entanglement productions, which is measured by the
linear entropy of the subsystem, with the help of the perturbative
formula, Eq.~(\ref{eq:miyaji}).
With this in mind, we discuss the recent study by Miller and
Sarkar~\cite{MS99}, who investigated the weakly chaotic region
(where chaotic seas and tori coexist) of the CKTs, and claimed that 
{\it the increment of the strength of chaos enhances entanglement}. More
precisely, they numerically found that the entanglement production
rate, which is measured by the von Neumann entropy, linearly depends
on the sum of positive (finite-time) Lyapunov exponents of the
corresponding classical CKTs, without any theoretical justification.
%
As is well known, it is much harder to develop a theory of weakly
chaotic systems (in other words, mixed phase space systems) than strongly
chaotic systems. This is actually the case with the numerical result
of Miller and Sarkar.
To accommodate these two qualitatively different results,
we employ our perturbative formula, Eq.~(\ref{eq:miyaji}), 
in the analysis for the weakly chaotic region ($k=3.0$, 
Fig.~\ref{fig:LsVsGweak}(a)).

In order to justify the application of 
our formula, Eq.~(\ref{eq:miyaji}), we confirm
that the entanglement productions measured by the linear entropy,
instead of the von Neumann entropy, reproduce Miller and Sarkar's
fitting. See Fig.~\ref{fig:LsVsGweak}(b). 
Furthermore, we numerically examined that the perturbative evaluation
of the entanglement production rate $\Gamma$, Eq.~(\ref{eq:miyaji}),
is applicable to the weakly chaotic regions, in particular the case
above. Hence, in the following, 
we reexamine the inputs of the 
formula, Eq.~(\ref{eq:miyaji}), 
which is the correlation functions of the uncoupled systems.

We focus on our assumption, Eq.~(\ref{eq:Dpheno}), for
the correlation  functions $D(l,m)$, Eq.~(\ref{eq:defD}), which is derived
from Eq.~(\ref{eq:correlation}) for strongly chaotic regions: 
(i) Due to the absence of tori in the corresponding classical
system, the fluctuation of $z_i$ takes a saturated value which agrees
with that of the uniform distribution in the classical phase space
(i.e., $C_i(l,l) \simeq \sigma_{\rm sat}^2$);
(ii) Due to the strongly chaotic dynamics, the correlation functions
decays exponentially (i.e. $C_i(l,m) \propto e^{-\gamma_i |l-m|}$).

\begin{figure}
\includegraphics[scale=0.9]{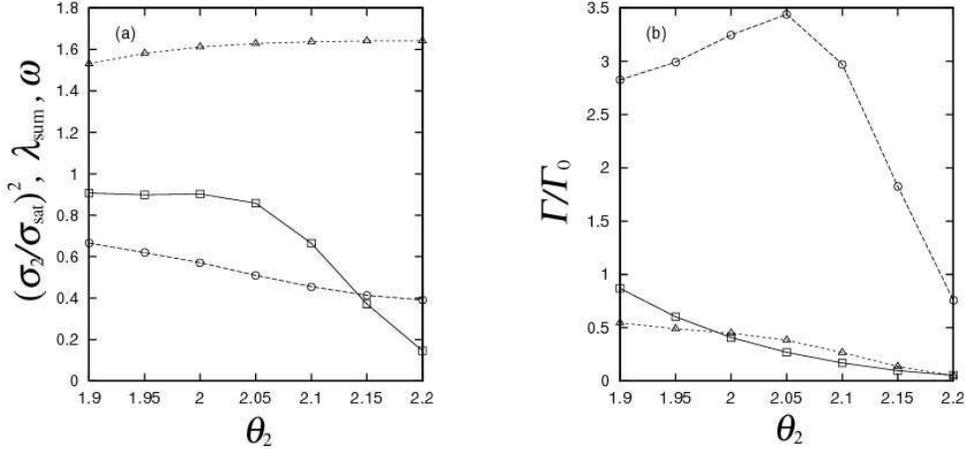}
\caption{\label{fig:T2VsGweak}
(a) $\theta_2$ dependences of 
$(\sigma_2/\sigma_{\rm sum})^2$ ($\square$), 
$\lambda_{\rm sum}$ ($\bigcirc$), 
and $\omega$ ($\triangle$). 
(b) The $\theta_2$ dependence of $\Gamma/\Gamma_0$ ($\square$).
The perturbative estimations with the assumptions~(\ref{eq:Dpheno})
and~(\ref{eq:impDpheno}) are indicated by $\bigcirc$ and $\triangle$,
respectively. Parameters and initial conditions are the same as in
Fig.~\ref{fig:LsVsGweak} (b). At the same time, we assume 
$({\sigma_1}/{\sigma_{\rm sat}})^2 = 1$ for the perturbative
estimations. 
}
\end{figure}

First, we examine our assumption that $\sigma_2$, the
fluctuation of $z_2=J_z/j$, is independent of $\theta_2$. In the
weakly chaotic region, this assumption breaks down due to the
confinement of phase-space dynamics by tori. Actually, $\sigma_2$
becomes smaller as the ``overlapping'' between the initial state and
tori become larger (see Fig.~\ref{fig:T2VsGweak}(a)).
By taking account of this fact into the
assumption Eq.~(\ref{eq:Dpheno}) on $D(l,m)$,
the decrement of $\sigma_2$ in tori provides a crudest explanation of
the $\theta_2$ dependence of $\Gamma$ 
(denoted by $\square$ in Fig.~\ref{fig:T2VsGweak}(b)).
That is, the decrement of the fluctuation of $z_2$ 
(denoted by $\square$ in Fig.~\ref{fig:T2VsGweak} (a)) 
due to the influence from tori
inhibits the entanglement production. 
See the line with $\square$ in Fig.~\ref{fig:T2VsGweak} (b).
However, the improved estimation denoted by 
$\bigcirc$ in Fig.~\ref{fig:T2VsGweak} (b)
still exhibits a qualitative discrepancy.

Second, to overcome this discrepancy, we improve the
assumption for $D(l,m)$ as
\begin{equation}
  \label{eq:impDpheno}
  D(l,m) =
  \sigma_{1}^2\sigma_{2}^2 e^{-\gamma |l-m|} e^{i\omega (l-m)},
\end{equation}
where we introduce a real-valued parameter $\omega$.
This characterizes oscillations due to the regular motion of the second
top. The resultant oscillation of $D(l,m)$ tends to reduce the
value of $\Gamma$ in Eq.~(\ref{eq:miyaji})
(cf. Eq.~(\ref{eq:dMdt}))
:
\begin{equation}
  \label{eq:impGamma}
  \Gamma = 
  \frac{%
    \left({\sigma_1}/{\sigma_{\rm sat}}\right)^2
    \left({\sigma_2}/{\sigma_{\rm sat}}\right)^2
  }{1 + \{\sin(\omega/2) / \sinh(\gamma/2)\}^2}
  \times \Gamma_0 \coth(\gamma/2).
  %
\end{equation}
We determine the value of $\omega$ in Eq.~(\ref{eq:impDpheno}) from the
Fourier transformation of $D(l,m)$ (see Fig.~\ref{fig:T2VsGweak}(a)).  
We depict the estimation Eq.~(\ref{eq:impGamma}) (denoted by $\triangle$) 
also in Fig.~\ref{fig:T2VsGweak} (b).
We conclude that the assumption Eq.~(\ref{eq:impDpheno}) provides a
satisfactory improvement of the evaluation of $\Gamma$ for
the weakly chaotic region that Miller and Sarkar investigated.

From our argument, it is seen that the contribution from tori
also play a role for the determination of the entanglement 
production rate via $\sigma_2$ and $\omega$.
Thus it is suggested that the linear dependence of $\Gamma$
with the sum of the positive Lyapunov exponents of the corresponding
classical system is not intrinsic for the weakly coupling region.

\section{Summary and outlook}
\label{sec:summary}

We have studied how the strength of chaos affects 
the production rate of quantum entanglement of
the coupled kicked tops (CKTs).
When the coupling constant $\epsilon$ is small enough, 
the entanglement productions obey the perturbative
formula, Eq.~(\ref{eq:miyaji}). 
When the classical counterpart exhibits chaotic
behavior, there appears a ``stationary'' entanglement
production regime where the entanglement production rate is
well-defined. 
In the strongly chaotic limit, 
where the correlation functions of the uncoupled tops decay exponentially
fast, the perturbative formula, Eq.~(\ref{eq:miyaji}), predicts
that the entanglement production rate saturates
to a certain value. Our numerical experiment confirmed this
prediction.
This is a unexpected result since
the previous works show that 
the chaotic dynamics promotes a larger amount of quantum entanglement
compared with the regular dynamics 
\cite{Adachi92,Tanaka96,SKO96,FNP98,Lakshminarayan01}.

Our perturbative argument of the strongly chaotic region depends only
on the two points: (i) the 
time-correlation function of the interaction Hamiltonian decays
exponentially; (ii) the phase space distributions of the corresponding
classical subsystems become quickly uniform before
the stationary entanglement production starts.
Hence we expect that our result also holds for a wide variety of
classically chaotic systems.

At the same time, we reexamined the weakly chaotic region which is
recently investigated by Miller and Sarkar, who showed numerically
that the entanglement production rate linearly depends on the sum of
the positive linear stability exponents~\cite{MS99}.
Our perturbative approach provides a theoretical way to explain their
result: The entanglement production rate is controlled by the
combination of the decay rate, the magnitude and the 
oscillation frequency of the time-correlation function of the interaction
Hamiltonian. It is hard to believe that these factors are generally
determined only by the Lyapunov exponents. 
Rather, it is natural to expect that the behavior of the correlation
function is strongly influenced by the existence of tori.

We point out that the investigation of dynamical
production of quantum entanglement has relevance with 
that of quantum fidelity (measured by an
overlapping integral of two states that are evolved by slightly
different Hamiltonians) \cite{Fidelity,Prosen:PRE-65-036208}. 
The decay of fidelity and the production of entanglement 
correspond to the quantum relaxations against static and
dynamic disturbances, respectively. 
When the disturbance is small
enough, the perturbative approach will describe the leading
(``linear'') response. We showed that this is the case for 
the dynamical production of quantum entanglement. On the other hand,
concerning the evaluations of  quantum fidelity, Prosen 
{\it et al.} reported the success of a perturbative 
approach~\cite{Prosen:PRE-65-036208}. Both works predict that
the strong chaos suppresses the quantum relaxations in flow systems.
Furthermore, Prosen {\it et al.} reported that their theoretical
prediction on the static disturbances agrees with their numerical
experiments~\cite{Prosen:PRE-65-036208}. 
The recent studies of quantum  
fidelity for nonperturbative regimes~\cite{NPTQF} will be applicable to
the studies on dynamical productions of quantum entanglement.
We believe that such an effort will be fruitful to investigate 
``quantum chaos'' in many degrees of freedom systems
(see, e.g., Refs.~\cite{TAI89,Znidaric:qph-0209145}).

Finally, we point out a possible application of our work to the
studies of realistic systems. 
In the investigations of chemical systems with large degrees of
freedom~\cite{SBPR96} including biological systems~\cite{WC01},
it is important to estimate the entanglement (decoherence) rate.
Most studies on this problem rely on the approaches using 
the master equations or the influence functional technique. 
However, these approaches have a serious difficulty in
practical applications to chemical reaction dynamics, since the
time-scale separation of the two constituents (``the system'' and
``the environment'') in the whole system often breaks down.
In contrast to this, our approach only assumes the weakness 
of the coupling between the subsystems, which dynamically causes
entanglement in the whole system. 
Hence it has an ability to cope with the breakdown of the the
time-scale separation.
We expect that our approach will provide a useful tool to
investigate chemical reaction dynamics.

\acknowledgments
One of the authors (H.F.) thanks Dr.\ H. Kamisaka 
for providing him the subroutine of the Jacobi polynomials,
and Dr.\ T. Takami, Dr.\ C. Zhu, Professor H. Nakamura, 
Professor S.\ Okazaki, Professor T. Konishi, Professor K. Nozaki, 
Dr.\ G.V. Mil'nikov, and Dr.\ S. Hayashi
for useful discussions and comments.
A.T. thanks Professor A. Shudo for useful conversations.


\begin{thebibliography}{99}

\bibitem{WZ82}
J.A.~Wheeler and W.H.~Zurek (editor), 
{\it Quantum Theory and Measurement}
(Princeton University Press, Princeton, 1982).

\bibitem{NC00}
M.A.~Nielsen and I.L.~Chuang,
{\it Quantum Computation and Quantum Information}
(Cambridge University Press, Cambridge, 2000).

\bibitem{ZP94}
W.H.~Zurek and J.P.~Paz, 
Phys. Rev. Lett. {\bf 72}, 2508 (1994);
G.~Casati and B.V.~Chirikov, ibid. {\bf 75}, 350 (1995);
W.H.~Zurek and J.P.~Paz, ibid. {\bf 75}, 351 (1995).

\bibitem{decoherence}
D.~Giulini, E.~Joos, C.~Keifer, J.~Kupsch, I.-O.~Stamatescu,
and H.D.~Zeh, {\it Decoherence and the Appearance of a Classical World
in Quantum Theory} (Springer, Berlin, 1996);
W.H.~Zurek, 
Physics Today, {\bf 44}, 36 (1991);
Prog. Theor. Phys. {\bf 89}, 281 (1993);
e-print quant-ph/0105127.


\bibitem{Guztwiller}
M.C.~Gutzwiller, 
{\it Chaos in Classical and Quantum Mechanics}
(Springer-Verlag, New York, 1990).


\bibitem{Adachi92}{%
  S.~Adachi, in {\em Proceedings of ISKIT '92},
  edited by I.~Tsuda and K.~Takahashi 
  (ISIP, Iizuka, 1992), p.~76.}

\bibitem{Tanaka96}
A.~Tanaka, J. Phys. A: Math. Gen. {\bf 29}, 5475 (1996).

\bibitem{SKO96}
M.~Sakagami, H.~Kubotani, and T.~Okamura,
Prog. Theor. Phys. {\bf 95}, 703 (1996).

\bibitem{FNP98}
  K.~Furuya, M.C.~Nemes, and G.Q.~Pellegrino,
  Phys. Rev. Lett. {\bf 80}, 5524 (1998).

\bibitem{Lakshminarayan01}
  A.~Lakshminarayan,
  Phys. Rev. E {\bf 64}, 036207 (2001).

\bibitem{AFNP99}
  R.M.~Angelo, K.~Furuya, M.C.~Nemes, and G.Q.~Pellegrino,
  Phys. Rev. E {\bf 60}, 5407 (1999).

\bibitem{MS99}
  P.A.~Miller and S.~Sarkar, 
  Phys. Rev. E {\bf 60}, 1542 (1999).


\bibitem{TFM02}
  For a brief account of this paper, see        
  A.~Tanaka, H.~Fujisaki, and T.~Miyadera,
  Phys. Rev. E {\bf 66}, 045201(R) (2002).

\bibitem{Haake00}
  F.~Haake, M.~Ku\'s, and R.~Scharf, 
  Z. Phys. B {\bf 65}, 381 (1987);
  F.~Haake, {\it Quantum Signatures of Chaos}, 
  2nd edition (Springer-Verlag, Berlin, 2000).


%
%
\bibitem{Sakurai94}
  J.J.~Sakurai,
  {\it Modern Quantum Mechanics},
  revised edition (Benjamin/Cummings, New York, 1994).

%
%
\bibitem{VMK88}
  D.A.~Varshalovich, A.N.~Moskalev, and V.K.~Khersonskii,
  {\it Quantum Theory of Angular Momentum} 
  (World Scientific, Singapore, 1988). 

\bibitem{spincoherent}
  J.M.~Radcliffe, J. Phys. A {\bf 4}, 3313 (1971);
  F.T.~Arecchi, E.~Courtens, and R.~Gilmore, 
  and H.~Thomas, Phys. Rev. A {\bf 6}, 2211 (1972).
\bibitem{EntanglementMeasure}{%
    S.M.~Barnett, and S.J.D.~Phoenix, Phys. Rev. A {\bf 40},
    2404 (1989).}

\bibitem{BK83}
G.P.~Berman and A.R.~Kolovsky,
Physica D {\bf 8}, 117 (1983);
D.L.~Shepelyansky,
ibid. {\bf 8}, 208 (1983).

\bibitem{Kubo1985}
     R. Kubo, M. Toda, and N. Hashitsume,
    {\it Statistical Physics II} (Springer-Verlag, Berlin, 1985).
\bibitem{Slichter}
  C.P.~Slichter,
  {\it Principles of Magnetic Resonance}
  (Harper \& Row, New York, 1963).   
 
\bibitem{CCIF79}
  G.~Casati, B.V.~Chirikov, F.M.~Izailev, and J.~Ford,
  in {\it Stochastic Behavior in Classical and 
    Quantum Hamiltonian Systems},
  ed. by G.~Casati and J.~Ford, Lecture Notes in Physics,
  Vol. 93 (Springer, Berlin, 1979).
\bibitem{TAI89}
  M.~Toda, S.~Adachi, and K.~Ikeda,
  Prog. Theor. Phys. Suppl. {\bf 98}, 323 (1989).

\bibitem{Fidelity}{%
    H.M.~Pastawski, P.R.~Levstein, and G.~Usaj, 
    Phys. Rev. Lett. {\bf 75}, 4310 (1995).
  }
\bibitem{Prosen:PRE-65-036208}{%
         T.~Prosen, Phys.\ Rev.~E {\bf 65}, 036208 (2002);
         T.~Prosen and M.~\v{Z}nidari\v{c},
         J.\ Phys.\ A: Math.\ Gen.\ {\bf 35}, 1455 (2002).
       }
\bibitem{NPTQF}{%
        F.M.~Cucchietti, C.H.~Lewenkopf, E.R.~Mucciolo,
        H.M.~Pastawski, and R.O.~Vallejos,
        Phys. Rev. E {\bf 65}, 046209 (2002)
        and references therein.
}
\bibitem{Znidaric:qph-0209145}{%
         M.~\v{Z}nidari\v{c} and T.~Prosen, 
         e-print quant-ph/0209145.
       }

\bibitem{SBPR96}
  B.\ J.\ Schwartz, E.\ R.\ Bittner, O.\ V.\ Prezhdo, and P.\ J.\ Rossky,
  J.\ Chem.\ Phys. {\bf 104}, 5942 (1996);
  O.V.~Prezhdo and P.J.~Rossky,
  Phys. Rev. Lett.\ {\bf 81}, 5294 (1998);
  S.\ Okazaki, Adv.\ Chem.\ Phys.\ {\bf 118}, 191 (2001).

\bibitem{WC01}
  A.\ Warshel and Z.\ T.\ Chu,
  J.\ Phys.\ Chem.\ B {\bf 105}, 9857 (2001);
  S.\ Hayashi, private communication. 

\end{thebibliography}
\end{document}